\documentclass[journal]{IEEEtran}

\usepackage{amssymb,amsmath,epsfig}
\usepackage{epstopdf}
\usepackage{cite}
\usepackage{soul}
\usepackage{xcolor}
\usepackage{tikz}
\usepackage{multirow}
\usepackage{hhline}
\usepackage{breqn}

\newcommand{\aA}{\mathbf{a}}

\newcommand{\fF}{\mathbf{f}}

\newcommand{\nN}{\mathbf{n}}
\newcommand{\rR}{\mathbf{r}}

\newcommand{\sS}{\mathbf{s}}

\newcommand{\wW}{\mathbf{w}}

\newcommand{\yY}{\mathbf{y}}

\newcommand{\Aa}{\mathbf{A}}

\newcommand{\Cc}{\mathbf{C}}

\newcommand{\Ff}{\mathbf{F}}

\newcommand{\Hh}{\mathbf{H}}

\newcommand{\Ss}{\mathbf{S}}

\newcommand{\Rr}{\mathbf{R}}

\newcommand{\Ww}{\mathbf{W}}

\newcommand{\Yy}{\mathbf{Y}}
\newcommand{\Zz}{\mathbf{Z}}

\DeclareMathAlphabet\mathbfcal{OMS}{cmsy}{b}{n}

\ifCLASSINFOpdf
  
\else
  
\fi

\begin{document}
\title{Deep-learning Image Reconstruction for Real-time Photoacoustic System}
%
%
%

\author{MinWoo~Kim, Geng-Shi Jeng, Ivan Pelivanov, Matthew O'Donnell 
\thanks{M. Kim, G. Jeng, I. Pelivanov and M. O'Donnell are with the uWAMIT Center in the Department of Bioengineering at the University of Washington, Seattle, WA, 98105 USA (e-mail: [mkim180, gsjeng, ivanp3, odonnel]@uw.edu).}
}


\maketitle

\begin{abstract}
Recent advances in photoacoustic (PA) imaging have enabled detailed images of microvascular structure and quantitative measurement of blood oxygenation or perfusion. Standard reconstruction methods for PA imaging are based on solving an inverse problem using appropriate signal and system models. For handheld scanners, however, the ill-posed conditions of limited detection view and bandwidth yield low image contrast and severe structure loss in most instances. In this paper, we propose a practical reconstruction method based on a deep convolutional neural network (CNN) to overcome those problems. It is designed for real-time clinical applications and trained by large-scale synthetic data mimicking typical microvessel networks. Experimental results using synthetic and real datasets confirm that the deep-learning approach provides superior reconstructions compared to conventional methods.     
\end{abstract}


%
\IEEEpeerreviewmaketitle

\section{Introduction}

\IEEEPARstart{R}{eal}-time integrated photoacoustic and ultrasound (PAUS) imaging is a promising approach to bring the molecular sensitivity of optical contrast mechanisms into clinical ultrasound (US) systems. Laser pulses transmitted into tissue induce light absorption from endogenous chromophores or exogenous contrast agents, which in turn launch acoustic waves according to the photoacoustic (PA) effect that can be used for imaging. We have recently developed a customized system for simultaneous PA and US imaging using interleaved techniques at fast scan rates \cite{jeng2019realTime}. One of the potential clinical applications is real-time, quantitative monitoring of blood oxygenation in the microvasculature \cite{attia2019review, schellenberg2018hand}. It requires not only multiple measurements at different optical wavelengths, but also high image quality to preserve microvascular topology.

Similar to US beamforming, PA reconstruction widely uses a traditional delay-and-sum (DAS) algorithm \cite{szabo2004diagnostic} for simplicity. However, the limited view and relatively narrow bandwidth of clinical US arrays greatly degrade image quality due to the ultra-broad bandwidth nature of PA signals \cite{han2018review}. This ill-posed problem causes structure loss, low contrast, and diverse artifacts making image interpretation difficult. 

To address these challenges, many groups have adopted reconstruction techniques from US or radar imaging to PA imaging. In particular, adaptive approaches such as a Minimum Variance (MV) method was developed to reduce off-axis signal and sidelobe artifacts by assigning apodization weights based on statistics \cite{park2008adaptiv2e, synnevag2009benefits}. Reconstruction using Delay Multiply and Sum (DMAS) methods can achieve higher image contrast by enhancing signal coherence nonlinearly \cite{kirchner2018signed, matrone2014delay}. Also, iterative techniques for the inverse reconstruction problem have been consistently developed for PA imaging using signal sparsity and low-rankness \cite{guo2010compressed, lin2017compressed}. All these methods may improve image quality by adopting sophisticated models based on system physics, data statistics, and underlying object properties. However, the main drawbacks are high computational complexity and the requirement of carefully selecting a handful of parameters.

Currently, a new category of reconstruction methods has been inspired by the field of machine learning (ML). Due to the great success of deep convolutional neural networks (CNN) in computer vision, reconstruction using supervised learning is an emerging research area in medical imaging  \cite{lee2017deep}. The network extracts best features via learning weights/filters to mitigate ill-posedness in inverse problems. The most popular ML framework uses learning in the image-domain, where training inputs are corrupted images processed by a standard reconstruction method under ill-posed conditions \cite{mccann2017review,cai2018end,antholzer2019deep, davoudi2019deep, awasthi2019pa}. Using this approach, the network avoids trying to capture detailed reconstruction operations and concentrates on filtering artifacts and noise. However, since the details of actual detector data are lost after image reconstruction, ML applied to reconstructed images often cannot recover weak signals and fine structures can be lost. 

Some frameworks are based on iterative schemes to train regularizations in Compressed Sensing (CS), but their extensive computations restrict real-time clinical applications \cite{boink2019partially,aggarwal2018modl}. Zhu et al proposed a framework starting with acquired data without prior knowledge of physics, but a fully connected layer requires a large number of weighting parameters for large data sets \cite{zhu2018image}. Allman et al employed PA raw data, but the application was only limited to the classification of point-like targets from artifacts \cite{allman2018photoacoustic}. 

Here we explore practical PA image reconstruction based on a deep-learning technique suitable for real-time PAUS imaging. We first examine the link between model-based methods and basic neural network layers to help design and interpret the learning structure. As discussed below, this study led us to modify 2-D raw data (with time and detector dimensions) into a 3-D array (with two spatial dimensions and a channel dimension), where a channel packet corresponds to the propagation delay profile for one spatial point, as an input to the neural network. The delay operation simplifies the learning process and the extension to channel dimension retains more information and increases learning accuracy. 

Our subsequent architecture is based on U-net \cite{ronneberger2015u}, where dyadic scale decomposition can access data in multi-resolution support. The structure can extract comprehensive features from the transformed 3-D array, replacing hand-crafted functions and generalizing standard filtering techniques. For training, we restrict the scope of absorber types to microvessels and create synthetic datasets using simulation. Operators transforming ground-truth to radio frequency (RF) array data are based on our current fast-swept PAUS system \cite{jeng2019realTime}, i.e. take into account the spectral bandwidth and geometry of a real imaging probe. However, it is not limited to only one imaging system and can be applied to any PA system with known geometry and characteristics.

To demonstrate the performance of the CNN-based method, we first compare it to standard methods using synthetic data. Then, we performed phantom experiments using the fast-swept PAUS system, and finally imaged a human finger in vivo.

\section{Signal and System Model}

\subsection{PA Forward Operation}

The spatiotemporal pressure change  $p(t,\rR)$ from short laser pulse excitation is captured in the photoacoustic equation \cite{wang2012biomedical},  
\begin{equation}\label{eq:PAFO}
\big(\nabla^2 - \frac{1}{v_s^2} \frac{\partial}{\partial t}\big) p(t,\rR) = -\frac{\beta}{C_p}\frac{\partial H(t,\rR)}{\partial t},
\end{equation}
where $v_s$ is the sound speed, $\beta$ is the thermal coefficient of volume expansion, and $C_p$ is the specific heat capacity at constant pressure. $H$ denotes the heating function given as $H=\mu_a \Phi$ where $\mu_a$ is the optical absorption coefficient and $\Phi$ is the fluence rate in a scattering medium. The heating function can be approximately described as $H(t,\rR) = s(\rR)\delta(t)$, where $s(\rR)$ denotes the spatial absorption function. The forward solution using the Green's function can be expressed as \cite{xia2014photoacoustic}
\begin{equation}\label{eq:GF}
p(t,\rR') = \frac{\Gamma }{4\pi v_s^2}  \frac{\partial}{\partial t} \bigg[ \int \frac{d\rR}{|\rR-\rR'|} s(\rR) \delta (t-\frac{|\rR-\rR'|}{v_s}) \bigg]
\end{equation}
where $\Gamma = \frac{\beta v_s^2}{C_p}$ is defined as the Grueneisen parameter and $\rR'$ is the detection position. Assume a transducer contains $J$ detection elements. 
Then, measurements  recorded by the $j$th element can be expressed as
\begin{equation}\label{eq:meas}
y(t,\rR'_j) = h(t)*\big[d(\rR'_j,\rR)p(t, \rR'_j)+ n(t, \rR'_j)\big] 
\end{equation}
where $d(\rR'_j,\rR)$, $h(t)$ and $n(\rR'_j,\rR)$ denote the directivity pattern, system impulse response and system noise, respectively, and $*$ represents the temporal convolution operator.  

\subsection{Limitations of Handheld Linear Array System}

\begin{figure}[t]
\centering
\includegraphics[width=8.7cm]{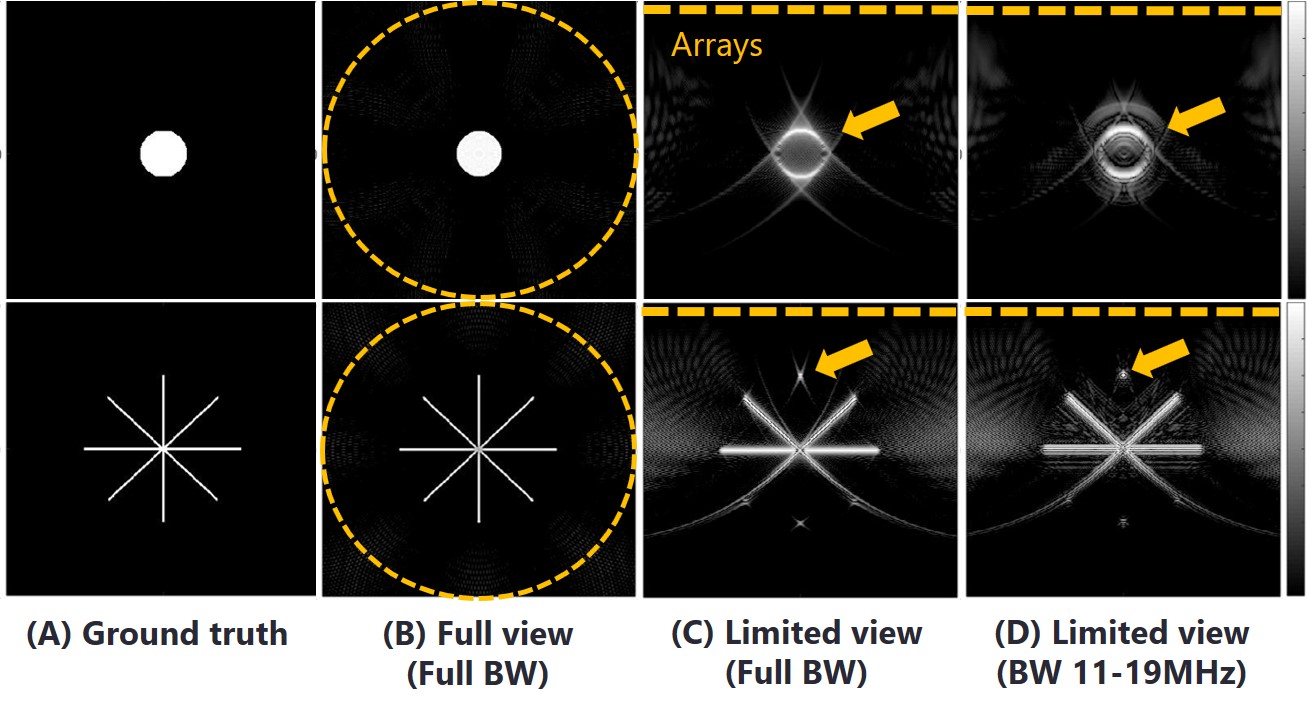}
\vspace*{-0.3cm}
\caption{Simulation results using standard filtered back-projection reconstruction. (A) presents two example object shapes. (B-D) shows reconstructions when the measurement conditions are (B) circular array with full bandwidth, (C) linear array with full bandwidth, and (D) linear array with limited bandwidth (11-19 MH). Arrows indicate structural losses and artifacts. All images are visualized on a log-scale colormap (40 dB range).   }
\label{fig:illPo}
\end{figure}

\begin{figure}[t]
\centering
\includegraphics[width=8.7cm]{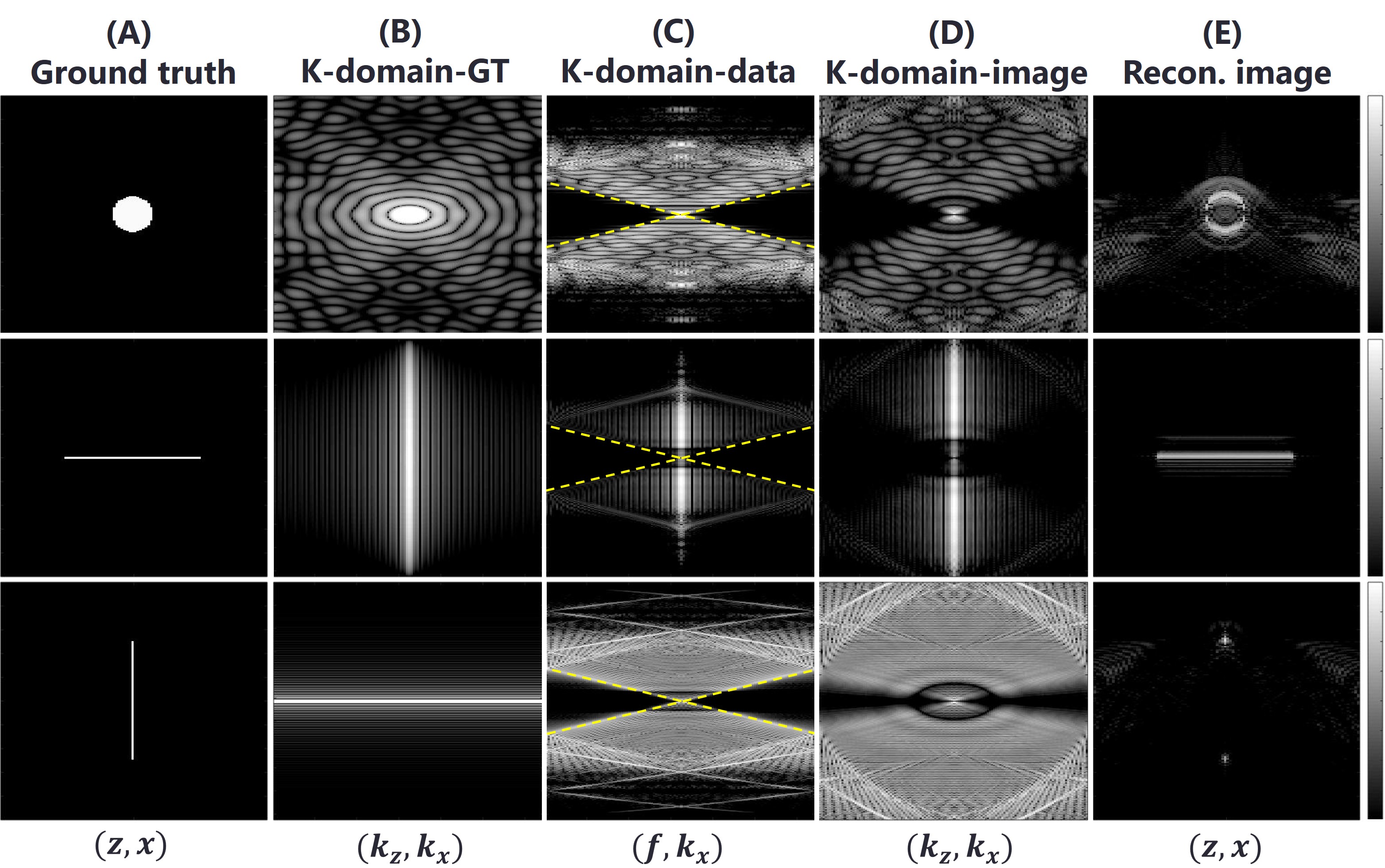}
\vspace*{-0.3cm}
\caption{Reconstruction simulations in k-space using one circular object and two simple linear objects rotated by 90 degrees. All images visualize absolute pixel values on a log-scale colormap (40 dB range). The maximum value in each image is represented as pure white.  (A) Ground-truth images. (B) K-domain-GT obtained by 2-D Fourier transforming the ground truth. (C) K-domain-data obtained by 2-D Fourier transforming raw data. Raw x-t data obtained with the forward model. The dotted lines indicate  $f=ck_x$. The empty region ($f<ck_x$) corresponds to evanescent waves. (D) K-domain-image by nonlinear mapping of K-domain-data. (E) Reconstructed image obtained by 2-D inverse Fourier transforming K-domain-image.  }
\label{fig:Simul}
\end{figure}

Reconstruction methods have been developed based on the measurement geometry. Filtered back-projection (FBP) is derived from the inverse of the PA forward operation in the spatiotemporal domain or k-space (frequency) domain \cite{xu2006photoacoustic}. Exact FBP formulas were demonstrated for a 3-D absorber distribution where the detection geometry is spherical, planar or cylindrical \cite{xu2017universal}. Imaging for a 2D spatial plane (slice) was adopted in standard tomographic scans assuming the detector (transducer) is focused in the plane \cite{wang2018finite, wang2019back}. A circular transducer for a 2D source distribution produces an accurate reconstruction if the number of detector elements provides enough spatial sampling density, as shown in Fig.~\ref{fig:illPo} (B) \cite{antholzer2019deep, xu2002time, haltmeier2013inversion, rosenthal2010fast}, and their bandwidth is not limited. In contrast, a linear sensor geometry  ($\rR'=x$) (as for a conventional US transducer) greatly limits the view, and also the bandwidth. Both limitations degrade image quality, as shown in Fig.~\ref{fig:illPo} (C) and (D). 

This ill-posed problem for a finite bandwidth linear array is more understandable in the frequency domain. For simplicity, assume that the directivity function is constant and the noise power is zero. Then, the 2-D Fourier transform of Eq. 3 for this geometry can be represented as \cite{xu2002exact}
\begin{equation}\label{eq:fourForm}
Y(f,k_x) = \alpha \frac{f sgn(f)}{\sqrt{(\frac{f}{c})^2-k_z^2}} S(k_z, k_x)
\end{equation}
where $k_z$ is defined by the mapping $k_z = sgn(f)\sqrt{(\frac{f}{c})^2-k_x^2}$.  As illustrated in Fig.~\ref{fig:Simul}, k-domain-data $Y(f, k_x)$ (Fig.~\ref{fig:Simul} (C)) are highly associated with the k-domain-image $S(k_z, k_x)$ (Fig.~\ref{fig:Simul} (D)) in spite of the nonlinear mapping. In addition to losses from evanescent waves, the narrow frequency bandwidth weakens low-frequency components of the object.  As shown in Fig.~\ref{fig:Simul} (E) (first row), reconstruction of a continuous absorbing medium is problematic for the linear array geometry.   

A special case of a vertical line source (bottom line in Fig.~\ref{fig:Simul}) exaggerates the problem. Its k-space spectrum is almost totally filtered, and only low amplitude spectral sidelobes invisible in the ground truth image survive. Thus, only top and bottom source points are visualized in the reconstruction. 

In this paper, the target objects of interest are microvessels, where the shape can be represented as a sum of straight and curvy lines. Since the signal components of the typical vascular structure are widely distributed in the k-domain, and a sufficient fraction are maintained even after limited view/bandwidth induced filtration, there is the possibility to reconstruct the entire object shape. However, note that it will be very challenging to recover vertical portions, as shown in Fig.~\ref{fig:illPo} and Fig.~\ref{fig:Simul}.

\section{Conventional Methods in PA Image Reconstruction}

\subsection{Delay and Sum (DAS)}

\begin{figure}[t]
\centering
\includegraphics[width=8.7cm]{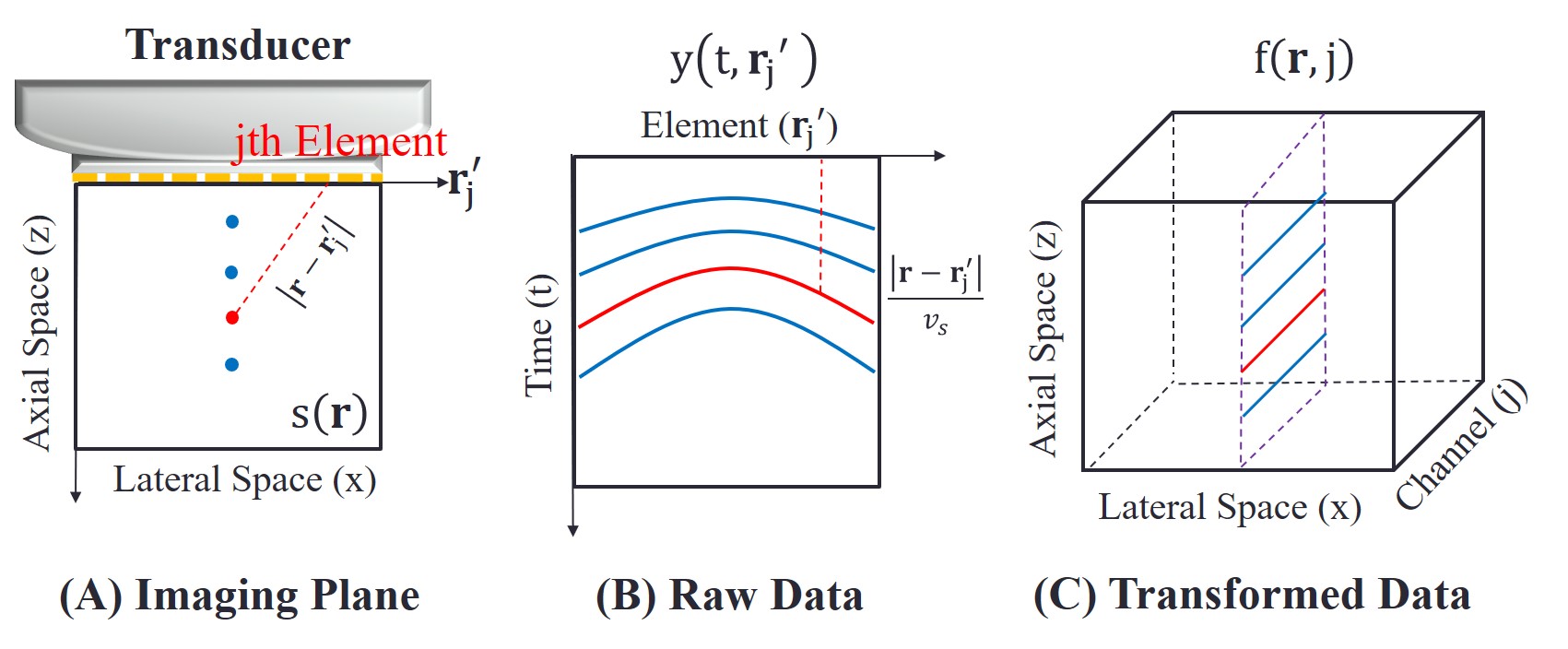}
\vspace*{-0.3cm}
\caption{(A) Measurement geometry. A 2-D image plane with respect to a linear array transducer is defined as the z-x plane. (B) 2-D measurement data. The curved lines indicate propagation delay profiles of particular image points at different depths. (C) 3-D transformed data. Channel packets correspond to the delay profiles indicated by straight lines.      }
\label{fig:transform}
\end{figure}

Most commercial US systems use DAS beamforming \cite{szabo2004diagnostic} for real-time image reconstruction. This procedure applies a delay due to the propagation distances between an observation point in the image and each transducer element prior to summation of signals across all array elements. Likewise, since the PA signal is based on one-way acoustic propagation, the DAS method can be applied as  
\begin{equation}
\tilde{s} (\rR) = \sum_j  w(\rR, \rR'_j) y (\frac{|\rR-\rR'_j|}{v_s},\rR'_j) 
\end{equation}
where $w(\rR, \rR'_j)$ denotes apodization weights. The framework of FBP is identical to DAS because DAS is associated with the adjoint of the forward operation (See Appendix I). Typically, standard DAS imaging applies a Hilbert transform after summation \cite{szabo2004diagnostic}. The expression can simply be represented by transforming data $y$ to $f$ as
\begin{equation} \label{eq:DASdisc}
\tilde{s} (\rR) = \sum_j  w(\rR, \rR'_j) f (\rR, j)
\end{equation}
where $f (\rR, j) = y (\frac{|\rR-\rR'_j|}{v_s}, \rR'_j)$. 
Fig.~\ref{fig:transform} illustrates the (delay) transformation provided that a detector is a linear-array transducer and the imaging plane is 2-D ($\rR=(z,x)$).  

\subsection{Minimum Variance (MV)}

MV is also based on Eq.~\ref{eq:DASdisc} but the weighting is adaptive \cite{park2008adaptiv2e}. For a position $\rR_i$, the vector form of Eq. ~\ref{eq:DASdisc} can be written as $\bar{s}=\bar{\wW}^T\fF_i$, where $T$ denotes transpose and vectors are in $R^{J}$. The weights are determined by minimizing the variance of $\bar{s}$ as
\begin{equation}
\min_{\bar{\wW}} \bar{\wW}^T E[\fF_i \fF_i^T] \bar{\wW} \;\; s.t. \;\; \bar{\wW}^T \bold{1} = 1,
\end{equation}
where $E[\cdot]$ denotes the expectation operator and the constraint forces unity gain at the focal point. The solution of the optimization problem is given as  
\begin{equation}
\bar{\wW} = \frac {\Rr_{i}^{-1} \bold{1}}{\bold{1}^H \Rr_{i}^{-1} \bold{1}}
\end{equation}
where $\Rr_{i} = E[\fF_{i} \fF_{i}^T]$.  
The covariance matrix  $\Rr_i$ can be estimated as 
\begin{equation}
\hat{\Rr}_i = \frac{1}{(2N+1)(J-L)} \sum_{n=-N}^{N}\sum_{l=0}^{J-L} \bar{\fF}_{i-n}^{(l)}\bar{\fF}_{i-n}^{(l) T} 
\end{equation}
where $\bar{\fF}_{i}^{(l)}=[\fF[1+l],\fF[2+l], \dots, \fF[L+l]]^T$ is the subarray for the position $\rR_i$, $L$ is the subarray length, and $2N+1$ is the number of averages over axial samples near the position $\rR_i$.   
Details can be found in \cite{park2008adaptiv2e}.  


\subsection{Delay Multiply and Sum (DMAS)}

The multi-channel array $f (\rR, j)$ is the beginning of the DMAS algorithm, as illustrated in Fig.~\ref{fig:standAlg} (B). Before summation, signal samples over channels at each position $\rR_i$ are combinatorially coupled as
\begin{align}
\tilde{s}(\rR_i) = \sum_{j_1,j_2} sgn(\bar{f}_{j_1,j_2}(\rR_i))\sqrt{|\bar{f}_{j_1,j_2}(\rR_i)|} 
\end{align}
where  $\bar{f}_{j_1,j_2}(\rR_i) = f(\rR_i, j_1)f(\rR_i, j_2)$.  
This nonlinear computation acts as a spatial cross-correlation, enhancing coherent signals while suppressing off-axis interference. The operations, sign and square root, are normalization steps to conserve signal power. The signal $\bar{f}_{j_1,j_2}(\rR_i)$ has modulated components near zero-frequency and harmonic components due to the coupling operation \cite{matrone2014delay}. Therefore, bandpass/highpass filtering is required in post-processing to suppress the components near zero-frequency. 

\subsection{Iterative Method with Compressed Sensing (CS)}

Iterative methods attempt to solve the inverse problem by adding regularization to overcome the ill-posed condition  \cite{guo2010compressed, lin2017compressed}. The matrix form of Eq.~\ref{eq:meas} can be expressed as $\yY = \Hh \sS + \nN$, where the matrix $\Hh$ is involved with the forward operation, directivity and system impulse response. The standard form of the inverse problem is given as 
\begin{equation}
\bar{\sS} = arg\min_{\sS} {||\yY-\Hh\sS||_2^2 + \lambda ||\Ww^T\sS||_1 }
\end{equation}
where $\Ww$ is the transform for sparsity, $\aA=\Ww^T\sS$ is the corresponding coefficient, and $\lambda$ is the regularization parameter.  Since the non-linear $l_1$ term does not allow a closed form solution, it is solved by iterative methods such as iterative shrinkage-thresholding argorithms (ISTA) and alternating direction method of multipliers (ADMM)  \cite{jin2017deep, gregor2010learning, boyd2011distributed}. For example, ISTA solves the optimization problem as 
\begin{equation}
{\sS}_{k+1} = \Ww \Theta_{\lambda\tau} \big( (\Ww^T-\tau(\Hh \Ww)^T \Hh)\sS_k + \tau (\Hh \Ww)^T \yY \big) 
\end{equation}
where $\Theta_{\alpha}$ is the soft-thresholding operator with value $\alpha$ and $1/\tau$ is the Lipschitz constant. The solution is updated by repetitive operations including matrix multiplication, matrix addition and thresholding. 

 The main disadvantage of MV, DMAS and CS is the high computational complexity for real-time imaging even though modifications have been proposed to reduce the burden. The selection of statistical operators or feature bases is a crucial step in model-based schemes. If the selection does not agree with the inherent properties of PA data, imaging is inaccurate. As an alternative, deep-learning approaches can build optimal feature maps through training and provide a practical reconstruction framework due to fast computation.

\section{Deep-learning reconstruction}

Recently, researchers have begun to find connections between conventional model-based approaches and deep convolutional neural networks (CNN) for inverse problems \cite{mccann2017review, jin2017deep}. We have also explored these links to build an efficient CNN for PA imaging using limited view/bandwidth arrays so that the network takes full advantage of signal characteristics such as row-rank, coherence and sparsity during learning. 

\subsection{Preprocessing}

It is not clear that a CNN can reconstruct absorption structures directly from PA data. Given the data dimension, learning would be extremely complex since the network architecture must encode the underlying PA forward operation. A popular approach to reduce this burden is preprocessing raw data with simple DAS reconstruction and using the resultant rough images as training input \cite{cai2018end, antholzer2019deep}. The CNN can then focus on learning the characteristics of artifacts in input images.  However, since these images can lose detailed information on object structure, the CNN output would not be perfect \cite{zhu2018image}. 

Our strategy is to use the transformed 3-D data $f(\rR_i,j)$ illustrated in Fig.~\ref{fig:transform} as the network input. As shown in the previous section, this operation is the first step for most reconstruction methods since it is based on the simple physics of wave propagation. The array represents delayed signals, where the delay is the propagation time from position $\rR_i$ to element $j$. Delayed data have several advantages: 1) detailed information embedded in raw data  $y(t,\rR)$ is not lost; and 2) it accelerates learning efficiency because channel samples at $\rR_i$ focus on waves coming from position $\rR_i$. 

This preprocessing is associated with row rank and sparsity in CS. The spatial domains combining axial and lateral dimensions  $\{\rR_i = (z_i, x_i) \}$ naturally reduce the number of bases, either patch-based or non-local, required to capture the essential features of microvessels. In addition, data extension to the channel axis can increase coefficient sparsity. That is, this representation can potentially reduce the rank of the problem and help discard off-target signals that introduce clutter. 

Note that MV and DMAS methods access the coherence using channel-sample correlation to indirectly enhance low-rank (high-coherence) signals while suppressing high-rank (less-coherence) artifacts and noise. The success of these methods suggests that preprocessing would help the CNN to find optimal filter bases (weights) for regression using a realistic number of examples in the training set.

\subsection{Structure}

\subsubsection{Operation and notation}
$\aA = v(\Aa) \in R^{nm \times 1}$ denotes vectorization by stacking the columns of the matrix  $\Aa \in R^{n \times m}$. Inversely, $\Aa = v_{n,m}^{-1}(\aA) \in R^{n \times m}$ denotes the matrix formed from the vector.  $\eta(\cdot)$ denotes the rectified linear unit function.  $\bold{1}_{n,m} \in R^{n \times m}$ denotes the matrix with every entry equal to one.  

\begin{figure}[t]
\centering
\includegraphics[width=8.7cm]{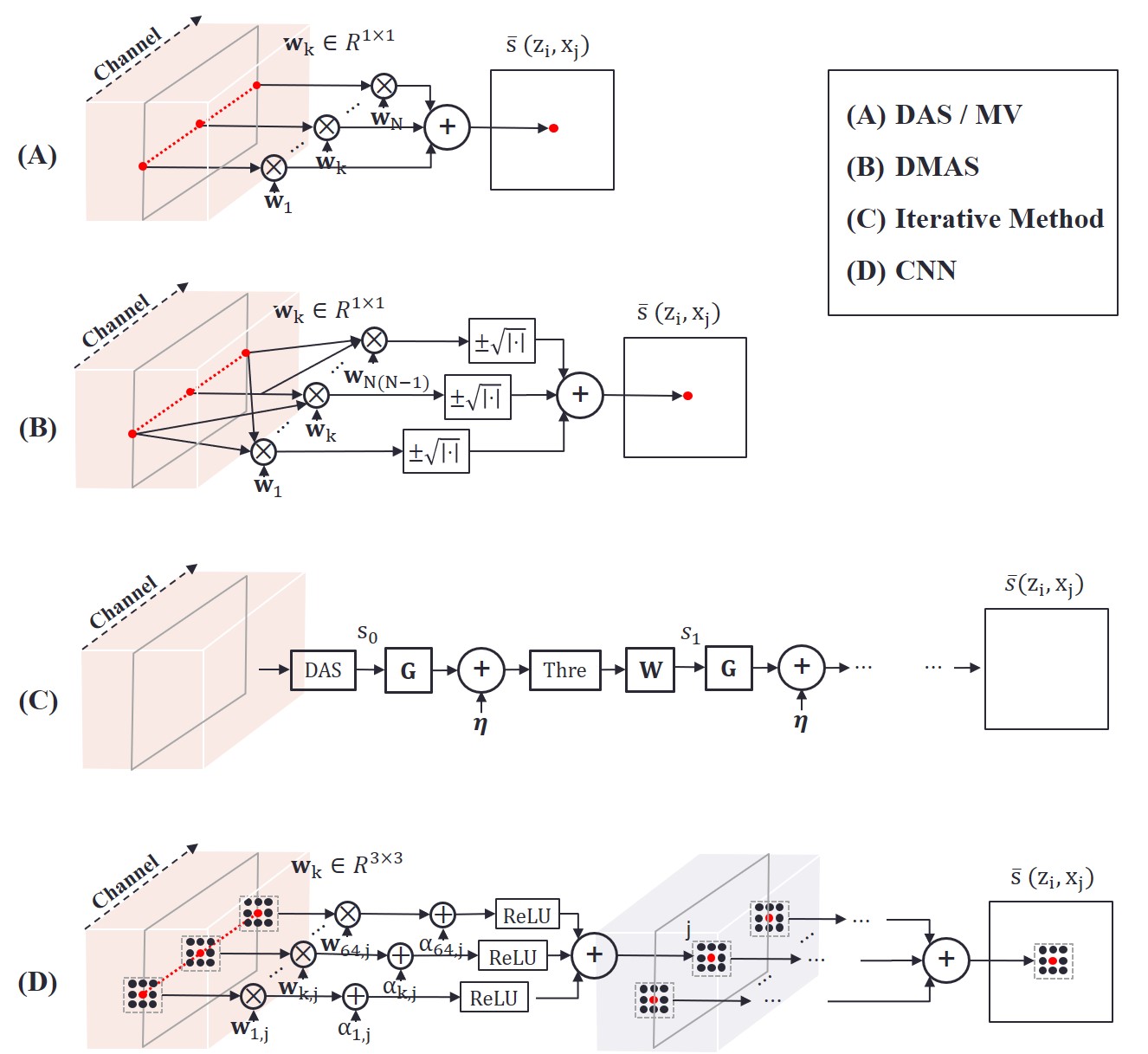}
\vspace*{-0.0cm}
\caption{Schematic diagram illustrating reconstruction methods. 2-D measurement data are transformed into a 3-D array as shown in  Fig.~\ref{fig:transform} followed by reconstruction. (A) DAS / MV methods. An image pixel is determined by weighting and summing channel samples at a corresponding pixel position. Weights vary with position. Unlike DAS, MV adaptively assigns weights depending on data statistics. (B) DMAS method. Channel samples are coupled and multiplied before summation. This additional nonlinear operation is required to prevent a dimensional problem. (C) Iterative method. This is based on the $L_1$ minimization problem in compressed sensing (CS). The initial solution is ordinarily obtained by DAS. The solution is updated by matrix multiplications, matrix additions, and threshold operations. (D) Basic structure of CNN. It applies convolution with a $3\times3$ kernel to multi-channel inputs, and returns multi-channel outputs. The full network consists of multiple layers, where each layer contains the convolution operation, bias addition and Rectified Linear Unit (ReLU) operation to enhance expressive power. }
\label{fig:standAlg}
\end{figure}

\begin{figure*}[tb]
\centering
\includegraphics[width=16cm]{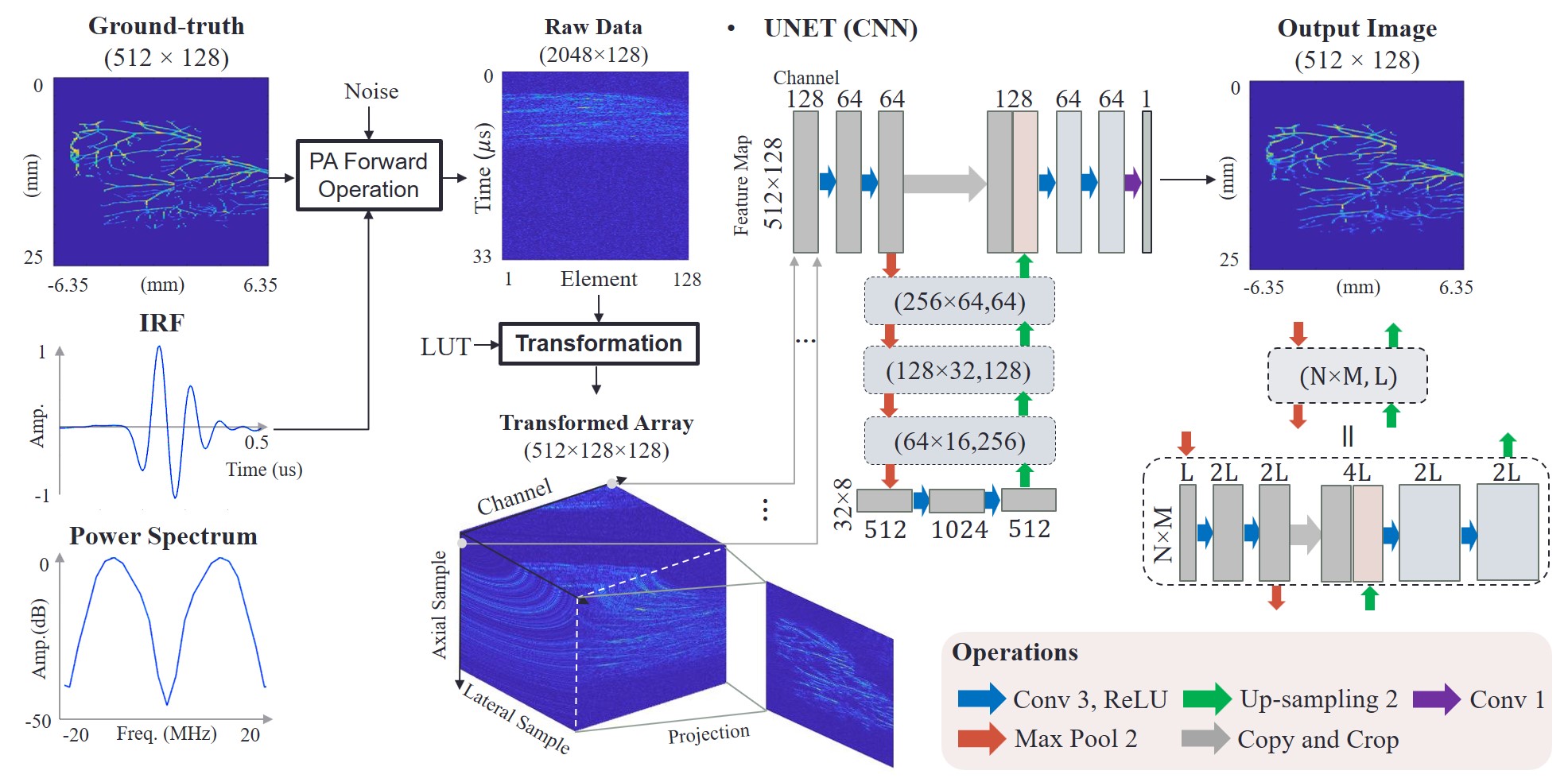}
\caption{Deep-learning architecture for PA image reconstruction. Raw data are converted into a 3-D array by a lookup table (LUT). The array is used as a multi-channel input to the network (first box). Each box represents a multi-channel feature map. The number of channels is denoted on the top or bottom of the box. Feature map sizes decrease and increase via max-pooling and upsampling, respectively. All convolutional layers consist of  $3 \times 3$ kernels except the last layer. The network is trained by minimizing the mean squared error between output images and ground-truth images.   }
\label{fig:unet}
\end{figure*}

\subsubsection{Encoder-decoder}
A standard encoder structure in CNN can be represented as  
\begin{equation}
v({\Cc_j}) = \bold{\Phi}^T \eta \bigg( \sum_i  v(\Ff_i  \circledast \Psi_{i,j} + \alpha_{j}\bold{1}_{n_1,n_2} ) \bigg)
\end{equation}
where $\Ff_i \in R^{n_1 \times n_2}$ is the $i$th channel of the input, $\Psi_{i,j}\in R^{d_1 \times d_2}$ are the learning weights for the $i$th channel of the input and $j$th channel of the output,  $\circledast$ is the 2-D convolution operation, $\alpha_j$ is the bias for the $j$th channel of the output, $\bold{\Phi}^T \in R^{m_1 m_2 \times n_1 n_2} $ is the pooling matrix, and $\Cc_j \in R^{m_1 \times m_2}$ is the $j$th channel of the encoding output.  
A corresponding decoder can be expressed as 
\begin{equation}
\Zz_j = \eta \bigg( \sum_i  v_{n_1,n_2}^{-1}\bigg(\big( \bar{\bold{\Phi}} v(\bar{\Cc}_i)  \big)\bigg) \circledast \Psi_{i,j} +  \beta_j \bold{1}_{n_1,n_2} \bigg)
\end{equation}
where $\bar{\Cc}_i \in R^{m_1 \times m_2}$ is the $i$th channel of the input, $\Psi_{i,j}\in R^{d_1 \times d_2}$ are the learning weights for the $i$th channel of the input and $j$th channel of the output, $\beta_j$ is the bias for the $j$th channel of the output, $\bar{\bold{\Phi}} \in R^{n_1 n_2 \times m_1 m_2}$ is the unpooling matrix, and $\Zz_j \in R^{n_1 \times n_2} $ is the $j$th channel of the decoding output.   

The encoder-decoder convolution layer is similar to the standard reconstruction methods shown in  Fig.~\ref{fig:standAlg}. The common structure is weighting (filtering) channel samples at a location $r_i$ or its neighborhood for an image pixel at $r_i$, whereas the scope of data locations (called the effective size or receptive field) contributing to a pixel varies with method. The methods based on DAS can assign different weights for every pixel. 

Although a CNN layer shares identical weights over space due to the convolution operation, the framework of multi-channel weights per layer and multi-layers increases the expressive power. The iterative method consists of matrix multiplications with no compacting support. While a CNN uses fixed filter size (usually $3\times3$), pooling operations enlarge the effective filter size in the middle layers.    

Currently, deep learning approaches have been investigated to understand the mathematical framework needed to solve inverse problems. Yin et al proposed the low-rank Hankel matrix approach using a combination of nonlocal basis and local basis for sparse representation of signals \cite{yin2017tale}. The framelet method has been successfully applied to image processing tasks since matrix decomposition reflects both local and nonlocal behavior of the signal. Ye et al discovered that the encoder-decoder framework of CNN generalizes the framelet representation \cite{ye2018deep}. In particular, the neural network can decompose the Hankel matrix of 3-D input data and shrink its rank to achieve a rank-deficient ground-truth (See Appendix B).

\subsubsection{Implementation Details} 

Our network is based on U-net. Fig.~\ref{fig:unet} presents the architecture. The left and right sides of the U-shape network correspond to successive encoders and decoders, respectively. Following a $3 \times 3$ convolutional layer and a ReLU layer, a batch normalization layer is used to improve learning speed and stability. The layers are repeated twice, either before  $2 \times 2$ max-pooling or after $2\times 2$ up-convolution (unpooling).  The pooling and unpooling operations allow multi-scale decomposition, where the size of feature maps are  $512\times128$, $256\times64$, $128\times32$ and $64\times16$. Total trainable parameters for all layers are 31,042,369.  
For fast preprocessing, we generated a sparse matrix (lookup table) mapping a 2-D raw data array $\Yy \in R^{2048 \times 128}$ into a 3-D delayed array $\Ff \in R^{512 \times 128 \times 128}$.    

\subsection{Network Training}
Training was performed by minimizing a loss function as
\begin{equation}
\hat{\Upsilon} = arg\min_{\Upsilon} {\sum_{k=1}^{K} ||\sS^{(k)}-\Upsilon(\Lambda^{(k)}, \Ff^{(k)})||_2^2  }
\end{equation} 
where $\Ff^{(k)}$ and $\sS^{(k)}$ denote the $k$th input and $k$th ground-truth, respectively, $K$(=16,000) denotes the total number of training datasets, and $\Upsilon$ denotes the trainable network structure. We exploited stochastic gradient descent (SGD) as an optimization method to minimize the loss function. The learning rate for SGD was set to 0.005 and the batch size is 8. 80$\%$ of total datasets were used for training and the rest were used for validation.  
The network was trained with a total of 150 epochs without over-fitting and under-fitting. To track the loss convergence of validation datasets, we defined a fractional error for the $j$th epoch as $\epsilon_j = |e_{j} - e_{j-1}|/e_{j} $ where $e_{j}$ and $e_{j-1}$ are the mean squared error (MSE) values at the $j$th and $j$-1th epoch, respectively. Fractional errors at the last 10 epochs are under 0.005. The MSE value at the final epoch for the validation datasets is 0.23, where the mean power value of the dataset is 1. Table~\ref{table1} summarizes the parameters.    

\begin{figure*}[tb]
\centering
\includegraphics[width=18.5cm]{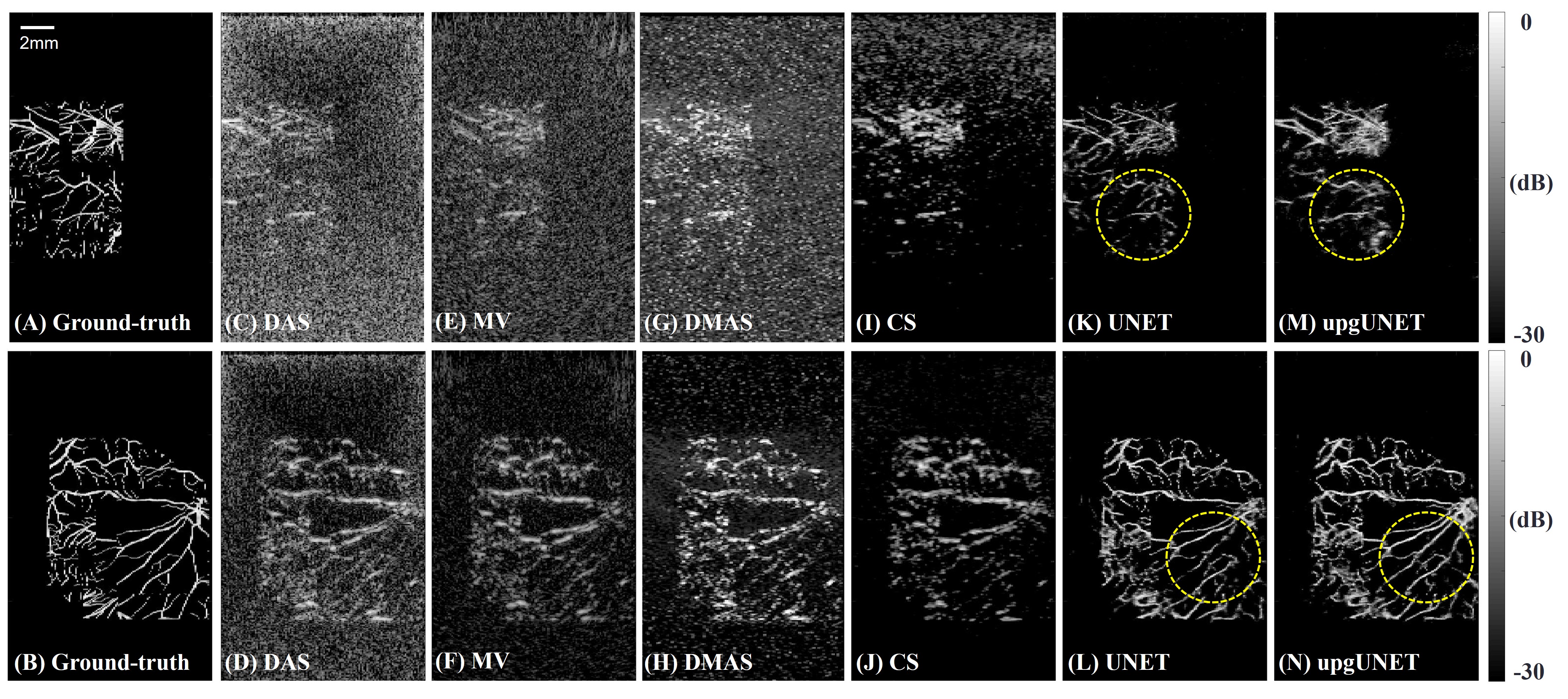}
\caption{Reconstruction results using synthetic data. Two particular objects are tested and all images are displayed using a log-scale colormap. (A,B) Ground-truth images. (C,D) Delay-and-sum results. Hilbert transform is applied as post-processing.  (E,F) Minimum variance results. (G,H) Delay-multiply-and-sum results. (I,J) Iterative CS method results. Wavelet dictionaries and total-variation regularization are used for compressed sensing. (K,L) Deep-learning results. An input is a 2-D array using DAS. (M,N) Deep-learning results. An input is a 3-D multi-channel array.}  
\label{fig:synResult}
\end{figure*}

\subsection{Training Data}

The supervised learning framework requires data at a large scale. However, it is mostly impractical to obtain clinical raw data accompanied by real ground-truth vascular maps. Therefore, we trained the network by creating synthetic data mimicking typical microvessel networks, as shown in  Fig.~\ref{fig:unet}.

The simulation transforming ground-truth to RF array data is based completely on our PAUS system. The impulse response function  $h(t)$ in Eq.~\ref{eq:meas} was measured by the system with a point source target.   Fig.~\ref{fig:unet} shows the response function and its power spectrum. The directivity is modeled as 
 \begin{equation}
D(\theta_j) = \frac{\sin(\frac{\pi l}{\lambda} \sin \theta_j)}{\frac{\pi l}{\lambda} \sin \theta_j}, \;\; \theta_j = \tan^{-1} (\frac{x-x_j^{'}}{z}),
\end{equation}
where $l$ is the transducer element pitch, $\lambda$ is the ultrasonic wavelength and $\theta_j$ is the incident angle of a wave propagating from position $\rR=(z,x)$ to the $j$th transducer element $\rR'_j=x'_j$ \cite{hasegawa2015effect}.   

Reference vascular images were obtained from the fundus oculi drive \cite{staal2004ridge}. The database contains retina color images captured by camera that can be used for vessel extraction. We used only binary images (manually extracted images), where white pixels denote the segmentation of blood vessels.   These images were randomly partitioned, re-sized, rotated and combined with each other to augment the training numbers. Next, every binary image was modified to a gray-scale image where the dynamic range of the vessel signal intensity is 20 dB. Lastly, every image was amplified with different values to obtain measurements involving a wide range of SNR.  Table ~\ref{table1} summarizes all parameters. 

\begin{table}[h]
\centering
\caption{Parameters for training network}
\label{table1}
\begin{center}
\begin{tabular}{ |l|l|l| }
\hline
\multirow{1}{8em}{Category/Function}
& Parameter 	& Value/Range \\
\hline
\multirow{7}{8em}{Raw data}
& Temporal samples	& 2048 \\
& Temporal sampling rate 	& 62.5 MHz \\
& Transducer aperture size  	& 12.8 mm \\
& Transducer element pitch	 & 0.1 mm \\
& Transducer element numbers 	& 128 \\
& Transducer center frequency	& 15.63 MHz \\
& Ultrasonic wavelength & 0.1 mm\\
\hline
\multirow{7}{8em}{Training image}
& Image numbers  	& 16,000 \\
& Signal dynamic range	& 20 dB \\
& Ratio (max signal/noise std)  	& 10-35 dB \\
& Vascular diameter  	& 0.05-0.3 mm \\
& Mean power & 1 \\
& Axial samples & 512 (25.6 mm) \\
& Lateral samples & 128 (12.7 mm) \\
\hline
\multirow{4}{8em}{Training}
& Batch size	& 8 \\
& Epochs  	& 150 \\
& Learning rate  & 0.005\\
& Trainable parameter numbers  	& 31,042,369 \\
\hline

\end{tabular}
\end{center}
\end{table}

\begin{table}[h]
\centering
\caption{ Parameters for reconstruction methods}
\label{table:recon}
\begin{center}
\begin{tabular}{ |l|l|l| }
\hline
\multirow{1}{4em}{Category}
& Parameter 	& Value/Type \\
\hline
\multirow{1}{4em}{DAS}
& f-number	& 0.1 or 0.5 \\
\hline
\multirow{2}{4em}{MV}
& Element number ($J$) 	& 128 \\
& Subarray length ($L$) 	& 32 \\
& Axial average number (2$N$+1)	& 5 \\
\hline
\multirow{3}{4em}{DMAS}
& Filter cutoff	& 6 MHz \\
& Highpass filter type  	& 6th-order Butterworth \\
\hline
\multirow{3}{4em}{CS}
& TV regularization ($\lambda_1$)	& 0.02 \\
& Wavelet type ($\Ww$)  	& Daubechies 4 \\
& Wavelet  regularization ($\lambda_2$)  & 0.005 \\
\hline
\end{tabular}
\end{center}
\end{table}

\section{Simulation Results}

We tested the reconstruction methods using both simulation and experimental data. We compared CNN-based approaches with other common reconstruction methods including DAS, DMAS and/or CS. Here, we call a network using DAS results (without Hilbert transform) a single-channel input `UNET' and a network using 3-D transformed arrays a multi-channel input `upgUNET'. As shown in Fig.~\ref{fig:unet}, UNET learns in the image-domain (2D projection) while upgUNET learns in the data-domain (3D array). 
DAS employs a rectangular apodization function where the activated aperture size is determined by f-number (=0.5). For image display, it uses the Hilbert transform following summation.  The iterative method exploits total-variation (TV) regularization and wavelet transforms for sparsity dictionaries.  We used the general criteria that the iteration stops when the norm of the gradient is less than a threshold value. We checked that more iterations barely change image quality or metric values.  We empirically selected all parameter values minimizing the error between ground-truth and reconstruction results. Table~\ref{table:recon} summarizes all parameters for the reconstructions.

The same procedure was used to generate testing and training datasets. However, two sets were generated from independent objects for independent verification. Fig.~\ref{fig:synResult} shows imaging results using the selected reconstruction methods from two particular examples where object shapes and data SNR are totally different. As expected, standard DAS (f-number=0.5) followed by Hilbert transformation provides low-contrast, poor-resolution images. While MV, DMAS and CS improve contrast in general, they often suppress weak signals. CNN-based methods restore most of the vasculature with stronger contrast and higher resolution. For upgUNET processing, fine vessels are more clearly visible, as shown in the circled areas of  Fig.~\ref{fig:synResult} (K) and (L). Lost structures are mostly vertically-extended vessels because their signal power is extremely low. 

In addition to these qualitative comparisons, we quantified performance differences employing the peak-signal-to-noise ratio (PSNR) and structure similarity (SSIM) metrics. The PSNR is defined via the MSE as 
\begin{equation}
\eta_{PSNR} = 10\log_{10} \bigg(\frac{n_1 n_2 I_{max}^2}{ ||\sS-\bar{\sS}||_F^2} \bigg)
\end{equation}  
where $||\cdot||_F$ denotes the Frobenius norm, $n_1\times n_2$ denotes the image size, and $I_{max}$ is the dynamic range (this value is 1 in our experiments). The SSIM is given as \cite{wang2004image}
\begin{equation}
\eta_{SSIM} = \frac{(2\mu_{\sS}\mu_{\bar{\sS}}+c_1)(2\sigma_{\sS,\bar{\sS}}+c_2) }{(\mu_{\sS}^2 + \mu_{\bar{\sS}}^2 + c_1)(\sigma_{\sS}^2 + \sigma_{\bar{\sS}}^2 + c_2) } 
\end{equation}  
where $\mu_{\sS}$, $\mu_{\bar{\sS}}$, $\sigma_{\sS}$ and $\sigma_{\bar{\sS}}$ denote the averages and standard deviations (i.e., square root of the variances) for $\sS$ and $\bar{\sS}$. $\sigma_{\sS,\bar{\sS}}$ denotes the covariance of $\sS$ and $\bar{\sS}$. Two variables $c_1=0.01^2$ and $c_2=0.03^2$, where used to stabilize the metric when either $(\mu_{\sS}^2 + \mu_{\bar{\sS}}^2)$ or $(\sigma_{\sS}^2 + \sigma_{\bar{\sS}}^2)$ is very close to zero. Since the resultant images have enough signal strength and deviation, the small variables are rarely influential. 

Table~\ref{table:Metric} presents average PSNR and SSIM values computed from 1,000 datasets where every vascular shape is different.
The numbers in parentheses are the standard deviations. 
Deep-learning approaches offer significant gain over standard methods, strongly suggesting that the network provides quantitatively better image quality. Note that the upgUNET produces the best values, in agreement with visual inspection.

\begin{table}[h]
\centering
\caption{Quantitative comparison of different methods}
\label{table:Metric}
\begin{tabular}{lllllll}
Metric 	&DAS	&MV		&DMAS	&CS		&UNET		&upgUNET	  	\\ \hline
PSNR		&20.97	&22.18	&22.32	&22.34	&26.71		&27.73		\\
(std.)		&(5.92)	&(4.72)	&(5.07)	&(4.70)	&(5.03)		&(5.21)		\\
SSIM		&0.208	&0.210	&0.260	&0.283	&0.745		&0.754		\\
(std.)		&(0.12)	&(0.11)	&(0.11)	&(0.14)	&(0.19)		&(0.20)		\\
\end{tabular}
\end{table}

\section{Experimental Results}

\subsection{PAUS System}

\begin{figure}[t] 
\centering
\includegraphics[width=8.7cm]{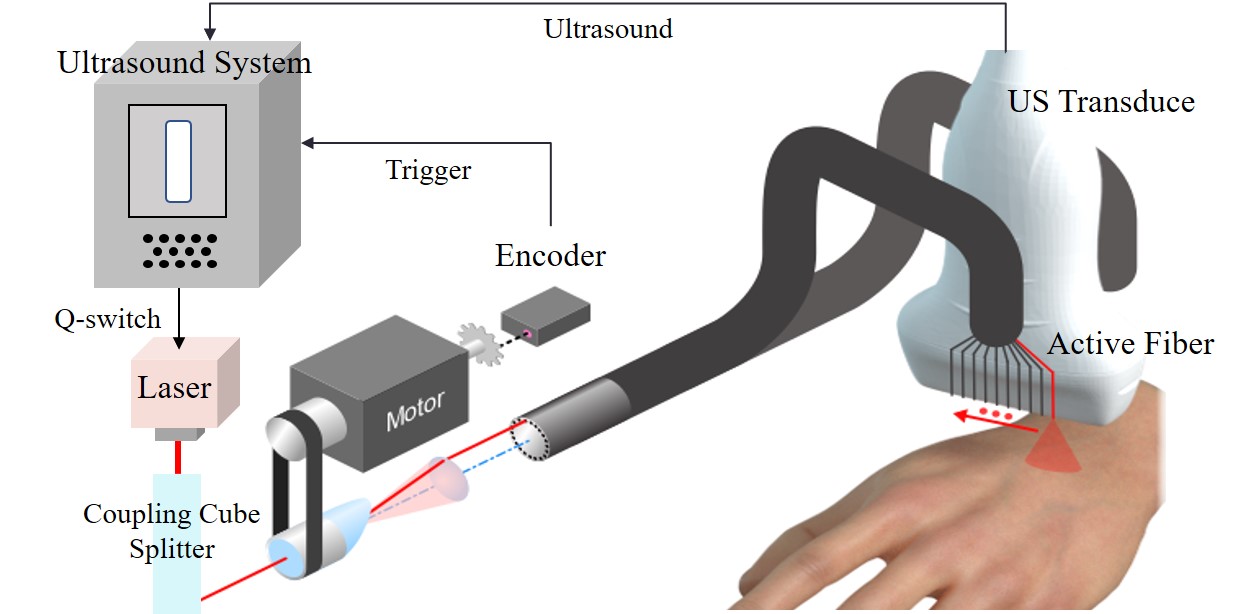}
\vspace*{-0cm}
\caption{Our customized PAUS system. An US scanner triggers a compact diode-pumped laser such that it emits pulses (around 1 mJ energy) at a 1 kHz rate with switching wavelength ranging from 700 nm to 900 nm. The laser is delivered to integrated fibers arranged on the two sides of a linear array transducer. A motor controlled by the scanner allows laser pulses to couple with different fibers sequentially. The scanner records PA signals that originate from light propagation into tissue.  }
\label{fig:PAUS}
\end{figure}

Our customized system for spectroscopic PA imaging is illustrated in  Fig.~\ref{fig:PAUS}. The scanner (Vantage, Verasonics, WA, USA) is programmed to record RF data at different wavelengths and fiber positions. A compact diode-pumped laser (TiSon GSA, Laser-export, Russia) transmits a light pulse at any arbitrary wavelength ranging from 700 to 900 nm at a 1 kHz rate. Transmitted pulses are sequentially delivered to 20 fiber terminals mounted around the top and bottom surfaces of a linear array transducer (LA 15/128-1633, Vermon S.A. France). The transducer center frequency is 15 MHz and the 3 dB bandwidth is around 8 MHz. US firings are interspersed with laser firings such that a full PA/US image frame at a fixed optical wavelength is recorded every 20 ms, producing a 50 Hz display rate for integrated US and PA images. System details are described in \cite{jeng2019realTime}.  
   
Here, for the purpose of reconstruction tests, we acquired data using one wavelength at 795 nanometers. The sampling rate for acoustic array data  $t_s$ is 62.5 MHz. The transducer contains 128 elements linearly arranged along the x-axis with a pitch of 0.1 mm. One PA data frame contains  2048 samples $\times$ 128 elements $\times$ 20 fibers.    
We averaged every data frame over fibers to enhance signal to noise ratio. The resultant data can be written as  $\Yy = y(t_k, \rR'_j)|_{t_k =kt_s} \in R^{2048 \times 128}$.  We reconstructed a 2-D image using each data frame. The image matrix can be expressed as  $\Ss = \bar{s}(\rR_i) = \bar{s}(z_i, x_i) |_{z_i = iz_r, x_i = ix_r} \in R^{512 \times 128}$ where axial and lateral resolutions are  $z_r=0.05$ mm and $x_r=0.1$ mm, respectively.   

\subsection{Phantom Study}

\begin{figure}[t]
\centering
\includegraphics[width=8.7cm]{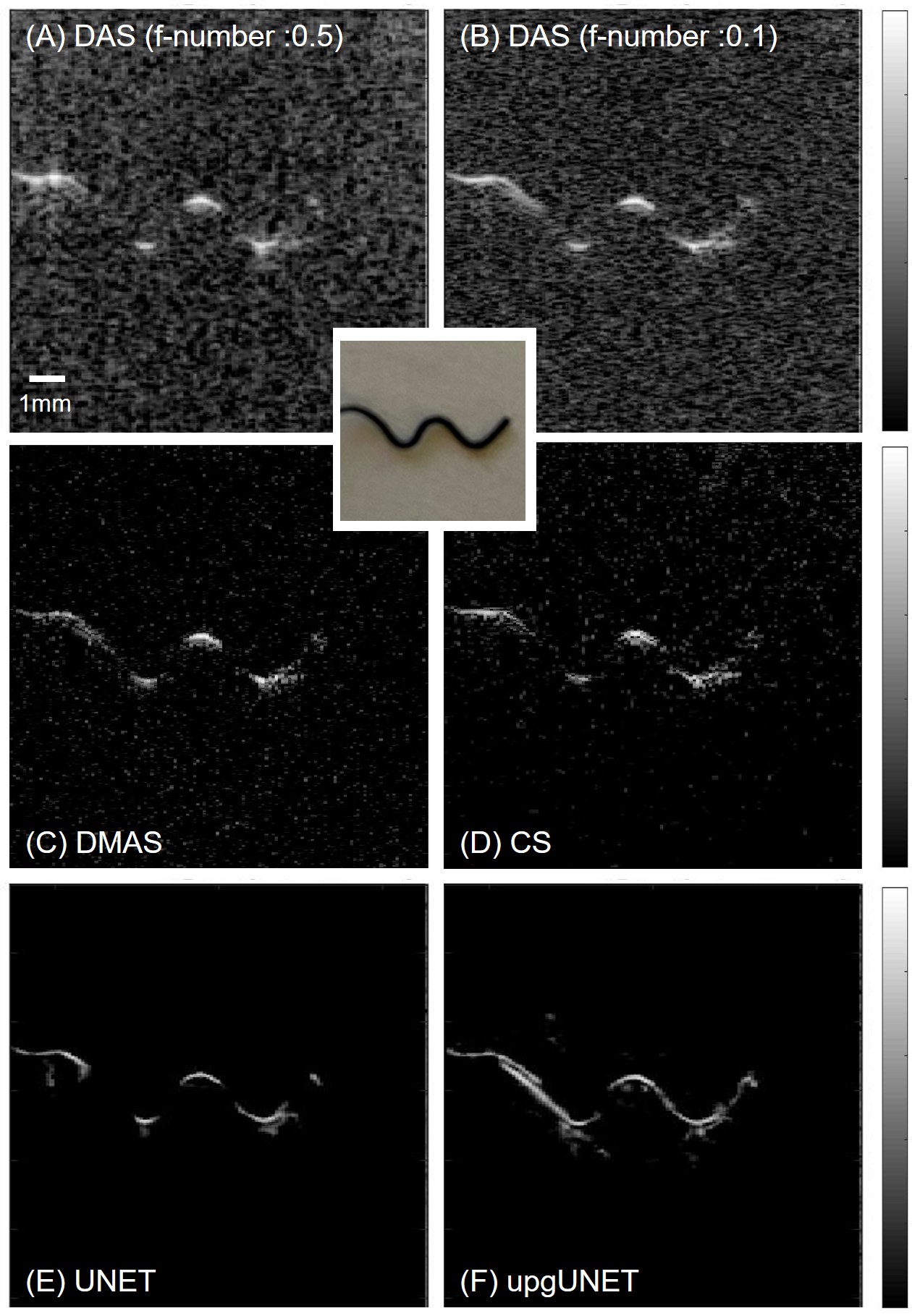}
\vspace*{-0.3cm}
\caption{Reconstruction results. All images are displayed using a log-scale colormap (35 dB range).  A `W' shape wire is scanned by the PAUS system. (A) Delay-and-sum result. The f-number is 0.5. (B) Delay-and-sum result. The f-number is 0.1. 
 (C) Delay-multiply-and-sum results. (D) Iterative CS method results. 
(E) UNET deep-learning result. An input is a 2-D array using DAS. (F) upgUNET deep-learning result. An input is a 3-D multi-channel tensor.} 
\label{fig:wshape}
\end{figure}  

\begin{figure}[t]
\centering
\includegraphics[width=8.7cm]{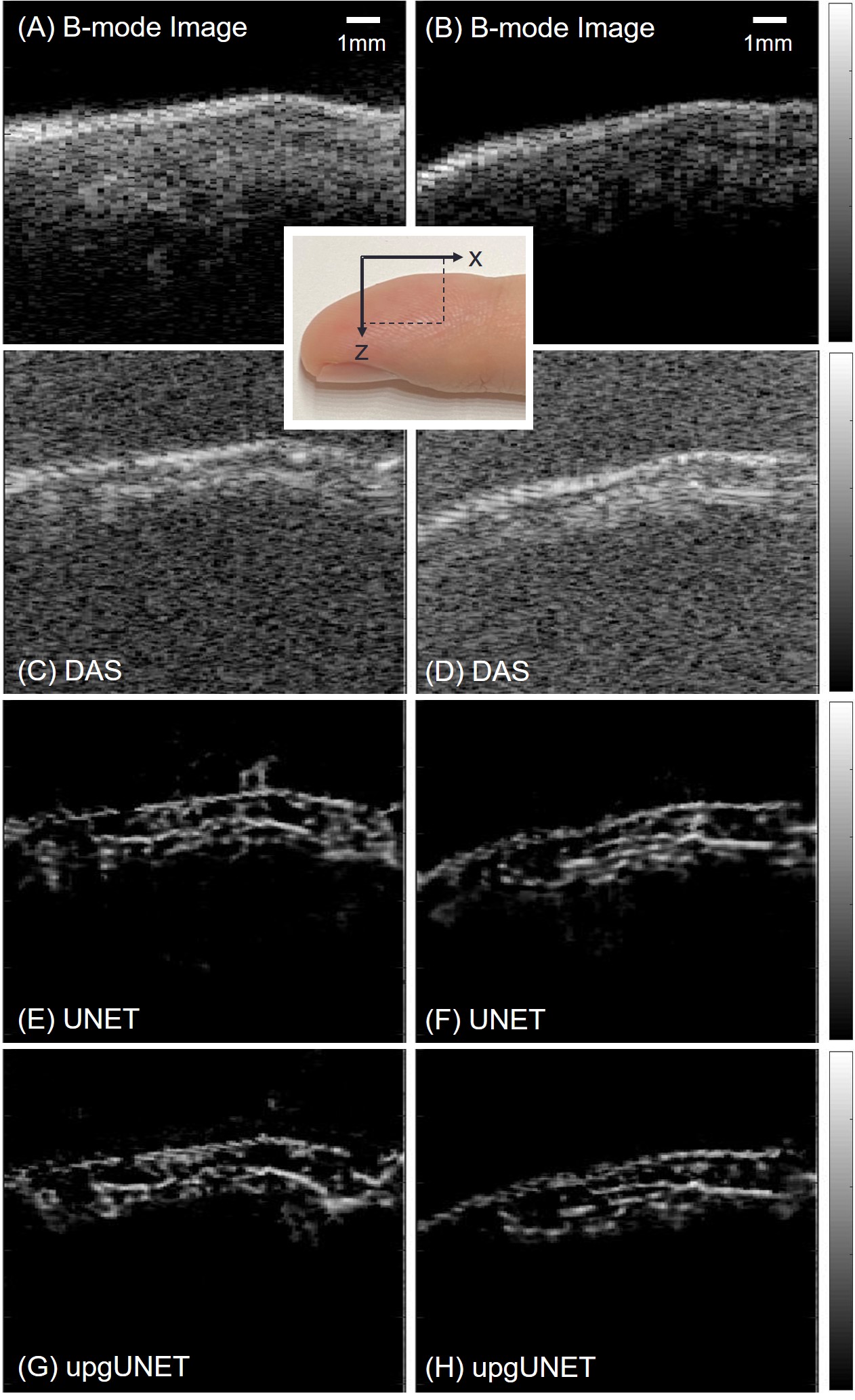}
\vspace*{-0.3cm}
\caption{ In vivo reconstruction results. A human finger is scanned by the PAUS system. Two sagittal planes are tested. (A, B) US B-mode image. (C, D) Delay-and-sum result. The f-number is 0.5. (E, F) UNET deep-learning result. An input is a 2-D array using DAS. (G, H) upgUNET deep-learning result. An input is a 3-D multi-channel tensor.  All images are displayed using a log-scale colormap. Mapping ranges for US and PD images are 50dB and 40dB, respectively. }
\label{fig:finger}
\end{figure}  

We constructed a phantom containing a metal wire acting as an optical absorption target. As shown in Fig.~\ref{fig:wshape}, the wire shape approximates the letter `W'. It was suspended from a cubical container such that it appears as the `W' shape in the z-x imaging plane. The container was filled with an intralipid solution (Fresenius Kabi, Deerfield, USA) acting as a scattering medium. The concentration of the intralipid is around 2$\%$ and the effective attenuation coefficient is around 0.1 mm$^{-1}$. Channel data were recorded with our customized system. 
Reconstruction results are shown in Fig.~\ref{fig:wshape}, which compares deep-learning methods with standard methods. For DAS, we tested a small f-number (0.1) in addition to the default value (0.5). The lower number corresponds to larger aperture size, which can access some information from diagonal lines at the expense of SNR in the entire field. 
In agreement with simulation results, DMAS and CS improve contrast but suppress weak structures. 

 In the deep-learning imaging results, object shapes are more distinct with higher resolution. In particular, the preferred upgUNET method restores most wire structure. One flaw in the deep-learning approaches is that the networks sometimes produce artifacts near objects, as shown in Fig.~\ref{fig:wshape} (C) and (D). We believe they arise from low-level reverberations not modeled in synthetic training data.

\subsection{In-vivo Test}

Lastly, we scanned and imaged a human finger to study the feasibility of our suggested method for in-vivo vascular imaging in real-time. These studies were approved by the Institutional Review Board (IRB) of the University of Washington (Study$\#$ 00009196) and used both optical and acoustic energies well within ANSI (optical) and FDA (US) guidelines. The PAUS system recorded US and PA measurements using an interleaved pulse sequence for 50 Hz frame-rate imaging. 

Fig.~\ref{fig:finger} (A) and (B) show two longitudinal cross-sections of the finger as US B-mode images. We applied DAS, UNET and upgUNET to PA reconstructions for visual comparison  (See Fig.~\ref{fig:finger} (C-H)). In this limited test, we found little difference between UNET and upgUNET reconstructions. We presume that upgUNET presents more realistic microvascular structures than B-mode images. However, it is obvious that deep learning images provide markedly superior contrast and resolution than equivalent images reconstructed with DAS. 

\section{Discussion}

As discussed in the introduction, the inverse problem of PA imaging can be solved exactly only when the detection surface represents a whole sphere, cylinder or infinite plane. These conditions can be obtained for small animal imaging  \cite{xia2013small} but are difficult to achieve in a clinical environment.

Hand-held US probes with relatively narrow bandwidth yields serious image quality losses for PA images, and such techniques alone will unlikely be accepted for medical use \cite{steinberg2019photoacoustic}. 
 
However interleaved spectroscopic PAUS has been recently shown to dramatically improve the capabilities of diagnostic US in monitoring interventional procedures even under limited view and bandwidth conditions \cite{schellenberg2018hand}. Very recently, a fast-sweep PAUS approach has been developed to operate at clinically acceptable frame rates for both PA and US modalities \cite{jeng2019realTime}.

The goal of this paper was to investigate whether deep-learning algorithms can improve the quality of images obtained with a hand-held US probe. We note that large-scale absorbing heterogeneities are unlikely to be fully corrected with the proposed upgUNET algorithm because the low frequencies associated with these objects are not preserved within the limited bandwidth of the detector. In our opinion, however, microvascular networks can be improved with advanced reconstruction algorithms based on upgNET. Although not fully validated over a wide range of experimental conditions, the simulations, phantom measurements and in vivo results presented here strongly support this statement. 
 
We have explored several different signal processing methods beyond traditional DAS reconstruction and found that deep learning has the potential for enhanced image quality and real-time implementation if input data are structured to match the network architecture of a reasonably sized net. Here we reformatted elemental signals from an US transducer array into multi-channel 3D data (tensor) using prior knowledge of propagation delays related to simple wave physics. Effective decomposition of this tensor by the neural network can capture latent structures and features.

Both experiments and simulations have demonstrated how the proposed neural network improves image quality for PA reconstruction. It produces images with stronger contrast, higher spatial resolution compared to DAS, and few structure losses even given the limited spatial and temporal bandwidth of the real-time system used for PA data acquisition. Additional image quality improvements were demonstrated using multi-channel data as the network input (upgUNET). The final advantage of this approach is that the network effectively learns filtering weights from training data while standard methods must explicitly impose parameter values, filtering dictionaries, or regularizers.  

Again, the novelty of our architecture is leveraging image reconstruction in the data-domain. It is less computational complex than end-to-end fully-connected networks that directly learn from raw-data. In addition, it preserves information better than image-to-image networks. Our architecture is based on U-net to efficiently cope with the non-stationary nature of off-axis signals, artifacts and noise, as well as to restore vascular structures. Compared to DL-based iterative schemes, it is much better suited to real-time imaging due to lower computations.

The computational cost of the matrix multiplication mapping a 2D data array  ($K \times J$) to a 3D array ($I_1 \times I_2 \times J$) is  O($KJ^2 I_1 I_2$). However, the operation matrix is mostly sparse, so the cost can be reduced as O($ \Gamma I_1 I_2 J$), where $\Gamma$ is the number of non-zeros per column. 
Convolution operations dominate computations in a CNN. The cost for an  $N \times N \times R_1$ input, $N \times N \times R_2$ output, $K \times K$ filters per layer and $L$ layers is $O(N^2 K^2 R_1 R_2 L)$. Since the operation is simple, it is ideal for parallel processing to reduce computational time. We employed Tensorflow with Keras to construct the network shown in  Fig.~\ref{fig:unet}. We implemented code for training and testing in Python and ran it on a computer using an Inter i7 and an NVIDIA 1080Ti. The code will be available on the website (https://github.com/bugee1727/PARecon) upon publication. 
  
The average computation time for image reconstruction from raw data is 40 msec, representing a 25 Hz real-time frame rate. Our current system functions at 50 Hz and higher frame rates, so these computation times must be reduced by at least a factor of two for true real-time implementation. Given an optimized hardware architecture for our specific reconstruction approach, a real-time frame rate of 50 Hz and higher is very realistic in the short term for the specific deep learning algorithm presented here for PA image reconstruction. 

The architecture combining a trainable neural network with a non-trainable preprocessor inputting raw data can be effective for diverse medical imaging fields and remote sensing applications. For example, some groups have recently investigated trainable apodization methods to replace slow MV beamforming for US B-mode imaging \cite{luijten2019adaptive, khan2019adaptive}. They showed that a simple network enables fast imaging computation without reducing MV image quality.  
     
One limitation of the CNN identified in these studies is poor image quality for structures aligned almost vertically (i.e., nearly parallel to the normal to the 1D transducer array). As expected, restoration is challenging for this case since most of the PA signal spatial frequencies are not captured by the transducer array. As shown in the top images of Fig.~\ref{fig:synResult}, this loss is more serious if the PA signal power of the object is weak compared with the noise power. Part of our future work will explore alternative approaches leveraging additional information from ultrasound imaging, or better system conditions such as acquiring one additional view at a significant angle with respect to the array normal at the first position. 

The second limitation of the method is the presence of some unexpected image artifacts for real data tests, which can reduce specificity. Obtaining real ground-truth maps is impracticable at large scale. Thus, reducing discrepancies between synthetic and real data is needed to guarantee that a trained CNN works best for real data. 

One of the main difficulties in reconstruction arises from the object dimension. Although a transducer lens focuses to a 2D plane (imaging plane), it cannot fully limit the sensitivity to a selected plane in an object. In other words, signals still come from points outside the imaging plane. Thus, generating training data based on 3D structures is required for the next step. Experiments are ongoing to map the full 3D PSF of our PA system and to include these details into the forward model for synthetic data generation. We expect that new synthetic data accompanied with an extended detection view will further improve image quality.  

There certainly are alternative learning frameworks. Our architecture is limited to supervised learning requiring ground-truth pairing with input data. Some architectures such as Generate Adversarial Networks (GANs) do not need pairs \cite{goodfellow2014generative, yi2019generative}. They can provide realistic fake images from measurements using real vascular image datasets obtained by other medical imaging modalities. Since only real data and real images are required for training, they can prevent unexpected artifacts. Also, different learning frameworks can bring additional information from US images to bear on PA image reconstruction.

Overall, our future studies will address these two limitations and develop specific deep learning architectures for real-time implementation. The goal is to create a CNN tuned to the problem of PA image reconstruction using limited spatial and temporal bandwidth data for robust, high-quality spectroscopic PAUS imaging at frame rates of 50 Hz and higher.

\section{Conclusions}
In this paper, we described a deep convolutional network for real-time PAUS imaging. The PAUS platform has the potential for real-time clinical implementation but the limited view/bandwidth of clinical US arrays degrades PA image quality. Our target of interest is microvessels, which can be almost entirely restored by advanced reconstruction methods, in contrast to large homogeneous objects. We developed a deep learning network for this application trained with realistic simulation data synthesized from ground-truth vascular image sets using a deterministic PA forward operator and the measured impulse response of a real transducer.
Reformatting raw channel data into a multi-channel array as a pre-processing step increased learning efficiency with respect to network complexity and imaging performance. The neural network is based on U-net, decomposing signals via multi-scale feature maps. Coupling the trainable network with the transformation method, we imaged structures mimicking vascular networks both in simulation and experiment. Overall, this approach reconstructs PA data with much higher image quality than conventional methods but loses some portions of complex absorber geometries and generates minor artifacts. 

\appendices
\section{Adjoint of operator $F$}
The solution of the forward model in Eq.~\ref{eq:GF} can simply be written as
\begin{align}
p(t,\rR') = F(s(\rR)), 
\end{align}
where $ \rR\in \Omega$ and $(t,\rR') \in (R \times \Xi)$.  The function $F(\cdot)$ is mapping $L_2 (\Omega) \rightarrow L_2 (R \times \Xi)$. Now, introduce $\tilde{p}(t,\rR') $ to define the adjoint of the operator $F$. The inner product between $F(s(\rR))$ and $q(t,\rR') $ can be expressed as
\begin{align}
&<F(s(\rR)), q(t,\rR') > \nonumber \\ 
&= \int_\Xi  \int_R \bigg [ \beta \int_\Omega \frac{\frac{\partial}{\partial t}\delta (t-\frac{|\rR-\rR'|}{v_s})}{|\rR-\rR'|} s (\rR) d\rR \bigg ] q (t,\rR') d\rR'dt   \nonumber \\
 &=  \int_\Omega s (\rR)  \bigg [ \int_\Xi  \frac{-\beta}{|\rR-\rR'|} \tilde{q} (\frac{|\rR-\rR'|}{v_s},\rR')  d\rR'   \bigg ] d\rR \nonumber \\
 &= <s, F^T(\tilde{q})> 
\end{align}
where $\beta = \frac{\Gamma }{4\pi v_s^2}$ and $\tilde{q} =\frac{\partial q}{\partial t}$. $F^T$ denotes the adjoint of the operator $F$. In general, DAS reconstruction ignores the derivative operation and applies a Hilbert transform after summation.

\section{Framelet Expansion}
Preprocessed data can be represented as a third-order tensor $\Ff \in R^{N \times M \times J}$ where $N$, $M$ and $J$ denotes the numbers of axial samples, lateral samples and channels. $\Ff_j = [\fF_{1,j},\dots,\fF_{m,j},\dots, \fF_{M,j}] \in R^{N \times M}$ denotes the matrix of the $j$th channel where $\fF_{m,j}$ denotes the $m$th column vector of the matrix. The block Hankel matrix for the periodic tensor is defined as
\begin{align}
H_{d_1,d_2}(\Ff) &= 
\begin{bmatrix} 
H_{d_1,d_2}(\Ff_1) & \cdots & H_{d_1,d_2}(\Ff_J)  
\end{bmatrix},  \nonumber \\
H_{d_1,d_2}(\Ff_j) &= 
\begin{bmatrix} 
H_{d_1}(\fF_{1,j}) & H_{d_1}(\fF_{2,j}) & \cdots & H_{d_1}(\fF_{d_2,j}) \\  
H_{d_1}(\fF_{2,j}) & H_{d_1}(\fF_{3,j}) & \cdots & H_{d_1}(\fF_{d_2+1,j}) \\
\vdots  &  \vdots   &   \ddots  &   \vdots  \\   
H_{d_1}(\fF_{M,j}) & H_{d_1}(\fF_{1,j}) & \cdots & H_{d_1}(\fF_{d_2-1,j}) 
\end{bmatrix},
\end{align}
where $H_{d_1,d_2}(\Ff) \in R^{NM \times d_1 d_2 J}$,  $H_{d_1,d_2}(\Ff_j) \in R^{NM \times d_1 d_2 }$.  The block matrix $H_{d_1}(\fF_{m,j})$ is given as  
\begin{align}
H_{d_1}(\fF_{m,j}) &= 
\begin{bmatrix} 
\fF_{m,j}[1] & \fF_{m,j}[2] & \cdots & \fF_{m,j}[d_1] \\  
\fF_{m,j}[2] & \fF_{m,j}[3] & \cdots & \fF_{m,j}[d_1+1] \\
\vdots  &  \vdots   &   \ddots  &   \vdots  \\   
\fF_{m,j}[N] & \fF_{m,j}[1] & \cdots & \fF_{m,j}[d_1-1] 
\end{bmatrix},
\end{align}
where $\fF_{m,j}[l]$ denotes the $l$th element of the vector. The Hankel matrix approach finds a solution $\hat{\Ff}$ minimizing $||\hat{\Ff}-\Ff^{*}||^2$ and the rank of the matrix $H_{d_1,d_2}(\hat{\Ff})$ where $\Ff^{*}$ denotes ground-truth. 
To address this minimization, a framelet approach handles the matrix decomposition using local and global bases as
\begin{equation}
H_{d_1 d_2} (\Ff) = \frac{1}{\alpha} \tilde{\bold{\Phi}}\bold{\Phi}^T H_{d_1 d_2} (\Ff) {\bold{\Psi}}\tilde{\bold{\Psi}}^T = \frac{1}{\alpha} \tilde{\bold{\Phi}}\bold{C}\tilde{\bold{\Psi}}^T
\end{equation}  
where $\tilde{\bold{\Phi}},\bold{\Phi} \in R^{N M \times p}$ denote non-local base pairs, $\tilde{\bold{\Psi}},\bold{\Psi} \in R^{d_1 d_2 J \times q}$ denote local base pairs, and $\bold{C}$ denotes framelet coefficients. This expression can be equivalently represented by an encoder-decoder structure if the reLU and bias are ignored for simplification. The non-local bases pairs correspond to pooling and unpooling operations. Thus, CNN can be interpreted as finding the best non-local bases and manipulating coefficients to access the minimization solution.


%



\ifCLASSOPTIONcaptionsoff
  \newpage
\fi



\bibliographystyle{IEEEtran}
\bibliography{IEEEabrv,IEEEexample}

\begin{thebibliography}{10}
\providecommand{\url}[1]{#1}
\csname url@samestyle\endcsname
\providecommand{\newblock}{\relax}
\providecommand{\bibinfo}[2]{#2}
\providecommand{\BIBentrySTDinterwordspacing}{\spaceskip=0pt\relax}
\providecommand{\BIBentryALTinterwordstretchfactor}{4}
\providecommand{\BIBentryALTinterwordspacing}{\spaceskip=\fontdimen2\font plus
\BIBentryALTinterwordstretchfactor\fontdimen3\font minus
  \fontdimen4\font\relax}
\providecommand{\BIBforeignlanguage}[2]{{%
\expandafter\ifx\csname l@#1\endcsname\relax
\typeout{** WARNING: IEEEtran.bst: No hyphenation pattern has been}%
\typeout{** loaded for the language `#1'. Using the pattern for}%
\typeout{** the default language instead.}%
\else
\language=\csname l@#1\endcsname
\fi
#2}}
\providecommand{\BIBdecl}{\relax}
\BIBdecl

\bibitem{jeng2019realTime}
G.-S. Jeng, M.-L. Li, M.~Kim, S.~J. Yoon, J.~J. Pitre, D.~S. Li, I.~Pelivanov,
  and M.~O'Donnell, ``Real-time spectroscopic photoacoustic/ultrasound (paus)
  scanning with simultaneous fluence compensation and motion correction for
  quantitative molecular imaging,'' \emph{bioRxiv}, 2019.

\bibitem{attia2019review}
A.~B.~E. Attia, G.~Balasundaram, M.~Moothanchery, U.~Dinish, R.~Bi,
  V.~Ntziachristos, and M.~Olivo, ``A review of clinical photoacoustic imaging:
  Current and future trends,'' \emph{Photoacoustics}, p. 100144, 2019.

\bibitem{schellenberg2018hand}
M.~W. Schellenberg and H.~K. Hunt, ``Hand-held optoacoustic imaging: A
  review,'' \emph{Photoacoustics}, vol.~11, pp. 14--27, 2018.

\bibitem{szabo2004diagnostic}
T.~L. Szabo, \emph{Diagnostic ultrasound imaging: inside out}.\hskip 1em plus
  0.5em minus 0.4em\relax Academic Press, 2004.

\bibitem{han2018review}
S.~H. Han, ``Review of photoacoustic imaging for imaging-guided spinal
  surgery,'' \emph{Neurospine}, vol.~15, no.~4, p. 306, 2018.

\bibitem{park2008adaptiv2e}
S.~Park, A.~B. Karpiouk, S.~R. Aglyamov, and S.~Y. Emelianov, ``Adaptive
  beamforming for photoacoustic imaging,'' \emph{Optics letters}, vol.~33,
  no.~12, pp. 1291--1293, 2008.

\bibitem{synnevag2009benefits}
J.-F. Synnevag, A.~Austeng, and S.~Holm, ``Benefits of minimum-variance
  beamforming in medical ultrasound imaging,'' \emph{IEEE transactions on
  ultrasonics, ferroelectrics, and frequency control}, vol.~56, no.~9, pp.
  1868--1879, 2009.

\bibitem{kirchner2018signed}
T.~Kirchner, F.~Sattler, J.~Gr{\"o}hl, and L.~Maier-Hein, ``Signed real-time
  delay multiply and sum beamforming for multispectral photoacoustic imaging,''
  \emph{Journal of Imaging}, vol.~4, no.~10, p. 121, 2018.

\bibitem{matrone2014delay}
G.~Matrone, A.~S. Savoia, G.~Caliano, and G.~Magenes, ``The delay multiply and
  sum beamforming algorithm in ultrasound b-mode medical imaging,'' \emph{IEEE
  transactions on medical imaging}, vol.~34, no.~4, pp. 940--949, 2014.

\bibitem{guo2010compressed}
Z.~Guo, C.~Li, L.~Song, and L.~V. Wang, ``Compressed sensing in photoacoustic
  tomography in vivo,'' \emph{Journal of biomedical optics}, vol.~15, no.~2, p.
  021311, 2010.

\bibitem{lin2017compressed}
X.~Lin, N.~Feng, Y.~Qu, D.~Chen, Y.~Shen, and M.~Sun, ``Compressed sensing in
  synthetic aperture photoacoustic tomography based on a linear-array
  ultrasound transducer,'' \emph{Chinese Optics Letters}, vol.~15, no.~10, p.
  101102, 2017.

\bibitem{lee2017deep}
J.-G. Lee, S.~Jun, Y.-W. Cho, H.~Lee, G.~B. Kim, J.~B. Seo, and N.~Kim, ``Deep
  learning in medical imaging: general overview,'' \emph{Korean journal of
  radiology}, vol.~18, no.~4, pp. 570--584, 2017.

\bibitem{mccann2017review}
M.~T. McCann, K.~H. Jin, and M.~Unser, ``A review of convolutional neural
  networks for inverse problems in imaging,'' \emph{arXiv preprint
  arXiv:1710.04011}, 2017.

\bibitem{cai2018end}
C.~Cai, K.~Deng, C.~Ma, and J.~Luo, ``End-to-end deep neural network for
  optical inversion in quantitative photoacoustic imaging,'' \emph{Optics
  letters}, vol.~43, no.~12, pp. 2752--2755, 2018.

\bibitem{antholzer2019deep}
S.~Antholzer, M.~Haltmeier, and J.~Schwab, ``Deep learning for photoacoustic
  tomography from sparse data,'' \emph{Inverse problems in science and
  engineering}, vol.~27, no.~7, pp. 987--1005, 2019.

\bibitem{davoudi2019deep}
N.~Davoudi, X.~L. De{\'a}n-Ben, and D.~Razansky, ``Deep learning optoacoustic
  tomography with sparse data,'' \emph{Nature Machine Intelligence}, vol.~1,
  no.~10, pp. 453--460, 2019.

\bibitem{awasthi2019pa}
N.~Awasthi, K.~R. Prabhakar, S.~K. Kalva, M.~Pramanik, R.~V. Babu, and P.~K.
  Yalavarthy, ``{PA}-fuse: deep supervised approach for the fusion of
  photoacoustic images with distinct reconstruction characteristics,''
  \emph{Biomedical optics express}, vol.~10, no.~5, pp. 2227--2243, 2019.

\bibitem{boink2019partially}
Y.~E. Boink, S.~Manohar, and C.~Brune, ``A partially learned algorithm for
  joint photoacoustic reconstruction and segmentation,'' \emph{arXiv preprint
  arXiv:1906.07499}, 2019.

\bibitem{aggarwal2018modl}
H.~K. Aggarwal, M.~P. Mani, and M.~Jacob, ``Modl: Model-based deep learning
  architecture for inverse problems,'' \emph{IEEE transactions on medical
  imaging}, vol.~38, no.~2, pp. 394--405, 2018.

\bibitem{zhu2018image}
B.~Zhu, J.~Z. Liu, S.~F. Cauley, B.~R. Rosen, and M.~S. Rosen, ``Image
  reconstruction by domain-transform manifold learning,'' \emph{Nature}, vol.
  555, no. 7697, p. 487, 2018.

\bibitem{allman2018photoacoustic}
D.~Allman, A.~Reiter, and M.~A.~L. Bell, ``Photoacoustic source detection and
  reflection artifact removal enabled by deep learning,'' \emph{IEEE
  transactions on medical imaging}, vol.~37, no.~6, pp. 1464--1477, 2018.

\bibitem{ronneberger2015u}
O.~Ronneberger, P.~Fischer, and T.~Brox, ``U-net: Convolutional networks for
  biomedical image segmentation,'' in \emph{International Conference on Medical
  image computing and computer-assisted intervention}.\hskip 1em plus 0.5em
  minus 0.4em\relax Springer, 2015, pp. 234--241.

\bibitem{wang2012biomedical}
L.~V. Wang and H.-i. Wu, \emph{Biomedical optics: principles and
  imaging}.\hskip 1em plus 0.5em minus 0.4em\relax John Wiley \& Sons, 2012.

\bibitem{xia2014photoacoustic}
J.~Xia, J.~Yao, and L.~V. Wang, ``Photoacoustic tomography: principles and
  advances,'' \emph{Electromagnetic waves (Cambridge, Mass.)}, vol. 147, p.~1,
  2014.

\bibitem{xu2006photoacoustic}
M.~Xu and L.~V. Wang, ``Photoacoustic imaging in biomedicine,'' \emph{Review of
  scientific instruments}, vol.~77, no.~4, p. 041101, 2006.

\bibitem{xu2017universal}
------, ``Universal back-projection algorithm for photoacoustic tomography,''
  in \emph{Photoacoustic Imaging and Spectroscopy}.\hskip 1em plus 0.5em minus
  0.4em\relax CRC Press, 2017, pp. 37--46.

\bibitem{wang2018finite}
B.~Wang, W.~Xiong, T.~Su, J.~Xiao, and K.~Peng, ``Finite-element reconstruction
  of 2d circular scanning photoacoustic tomography with detectors in far-field
  condition,'' \emph{Applied optics}, vol.~57, no.~30, pp. 9123--9128, 2018.

\bibitem{wang2019back}
B.~Wang, T.~Su, W.~Pang, N.~Wei, J.~Xiao, and K.~Peng, ``Back-projection
  algorithm in generalized form for circular-scanning-based photoacoustic
  tomography with improved tangential resolution,'' \emph{Quantitative imaging
  in medicine and surgery}, vol.~9, no.~3, p. 491, 2019.

\bibitem{xu2002time}
M.~Xu and L.~V. Wang, ``Time-domain reconstruction for thermoacoustic
  tomography in a spherical geometry,'' \emph{IEEE transactions on medical
  imaging}, vol.~21, no.~7, pp. 814--822, 2002.

\bibitem{haltmeier2013inversion}
M.~Haltmeier, ``Inversion of circular means and the wave equation on convex
  planar domains,'' \emph{Computers \& Mathematics with Applications}, vol.~65,
  no.~7, pp. 1025--1036, 2013.

\bibitem{rosenthal2010fast}
A.~Rosenthal, D.~Razansky, and V.~Ntziachristos, ``Fast semi-analytical
  model-based acoustic inversion for quantitative optoacoustic tomography,''
  \emph{IEEE transactions on medical imaging}, vol.~29, no.~6, pp. 1275--1285,
  2010.

\bibitem{xu2002exact}
Y.~Xu, D.~Feng, and L.~V. Wang, ``Exact frequency-domain reconstruction for
  thermoacoustic tomography. i. planar geometry,'' \emph{IEEE transactions on
  medical imaging}, vol.~21, no.~7, pp. 823--828, 2002.

\bibitem{jin2017deep}
K.~H. Jin, M.~T. McCann, E.~Froustey, and M.~Unser, ``Deep convolutional neural
  network for inverse problems in imaging,'' \emph{IEEE Transactions on Image
  Processing}, vol.~26, no.~9, pp. 4509--4522, 2017.

\bibitem{gregor2010learning}
K.~Gregor and Y.~LeCun, ``Learning fast approximations of sparse coding,'' in
  \emph{Proceedings of the 27th International Conference on International
  Conference on Machine Learning}, 2010, pp. 399--406.

\bibitem{boyd2011distributed}
S.~Boyd, N.~Parikh, E.~Chu, B.~Peleato, J.~Eckstein \emph{et~al.},
  ``Distributed optimization and statistical learning via the alternating
  direction method of multipliers,'' \emph{Foundations and
  Trends{\textregistered} in Machine learning}, vol.~3, no.~1, pp. 1--122,
  2011.

\bibitem{yin2017tale}
R.~Yin, T.~Gao, Y.~M. Lu, and I.~Daubechies, ``A tale of two bases:
  Local-nonlocal regularization on image patches with convolution framelets,''
  \emph{SIAM Journal on Imaging Sciences}, vol.~10, no.~2, pp. 711--750, 2017.

\bibitem{ye2018deep}
J.~C. Ye, Y.~Han, and E.~Cha, ``Deep convolutional framelets: A general deep
  learning framework for inverse problems,'' \emph{SIAM Journal on Imaging
  Sciences}, vol.~11, no.~2, pp. 991--1048, 2018.

\bibitem{hasegawa2015effect}
H.~Hasegawa and H.~Kanai, ``Effect of element directivity on adaptive
  beamforming applied to high-frame-rate ultrasound,'' \emph{IEEE transactions
  on ultrasonics, ferroelectrics, and frequency control}, vol.~62, no.~3, pp.
  511--523, 2015.

\bibitem{staal2004ridge}
J.~Staal, M.~D. Abr{\`a}moff, M.~Niemeijer, M.~A. Viergever, and
  B.~Van~Ginneken, ``Ridge-based vessel segmentation in color images of the
  retina,'' \emph{IEEE transactions on medical imaging}, vol.~23, no.~4, pp.
  501--509, 2004.

\bibitem{wang2004image}
Z.~Wang, A.~C. Bovik, H.~R. Sheikh, E.~P. Simoncelli \emph{et~al.}, ``Image
  quality assessment: from error visibility to structural similarity,''
  \emph{IEEE transactions on image processing}, vol.~13, no.~4, pp. 600--612,
  2004.

\bibitem{xia2013small}
J.~Xia and L.~V. Wang, ``Small-animal whole-body photoacoustic tomography: a
  review,'' \emph{IEEE Transactions on Biomedical Engineering}, vol.~61, no.~5,
  pp. 1380--1389, 2013.

\bibitem{steinberg2019photoacoustic}
I.~Steinberg, D.~M. Huland, O.~Vermesh, H.~E. Frostig, W.~S. Tummers, and S.~S.
  Gambhir, ``Photoacoustic clinical imaging,'' \emph{Photoacoustics}, 2019.

\bibitem{luijten2019adaptive}
B.~Luijten, R.~Cohen, F.~J. de~Bruijn, H.~A. Schmeitz, M.~Mischi, Y.~C. Eldar,
  and R.~J. van Sloun, ``Adaptive ultrasound beamforming using deep learning,''
  \emph{arXiv preprint arXiv:1909.10342}, 2019.

\bibitem{khan2019adaptive}
S.~Khan, J.~Huh, and J.~C. Ye, ``Adaptive and compressive beamforming using
  deep learning for medical ultrasound,'' \emph{arXiv preprint
  arXiv:1907.10257}, 2019.

\bibitem{goodfellow2014generative}
I.~Goodfellow, J.~Pouget-Abadie, M.~Mirza, B.~Xu, D.~Warde-Farley, S.~Ozair,
  A.~Courville, and Y.~Bengio, ``Generative adversarial nets,'' in
  \emph{Advances in neural information processing systems}, 2014, pp.
  2672--2680.

\bibitem{yi2019generative}
X.~Yi, E.~Walia, and P.~Babyn, ``Generative adversarial network in medical
  imaging: A review,'' \emph{Medical image analysis}, p. 101552, 2019.

\end{thebibliography}
\end{document}